\documentclass[12pt]{koval}

\usepackage{eucal}

\begin{document}

\title[Lyapunov exponents]{Lyapunov exponents in continuum Bernoulli-Anderson models}

\author{David Damanik}

\address{Department of Mathematics, University of California at Irvine, CA 92697-3875, USA}


\email{damanik@math.uci.edu}

\thanks{D.D.\ partially supported by the German Academic Exchange Service through HSP III
(Postdoktoranden)}

\author{Robert Sims}

\address{Department of Mathematics, University of Alabama at Birmingham, AL 35294-1170, USA}

\email{sims@math.uab.edu}
\thanks{R.S.\ partially supported by NSF Grant DMS-9706076}

\author{G\"unter Stolz}

\address{Department of Mathematics, University of Alabama at Birmingham, AL 35294-1170, USA}

\email{stolz@math.uab.edu}
\thanks{G.S.\ partially supported by NSF Grants DMS-9706076 and DMS-0070343}

\begin{abstract}
We study one-dimensional, continuum Bernoulli-Anderson models with
general single-site potentials and prove positivity of the
Lyapunov exponent away from a discrete set of critical energies.
The proof is based on F\"urstenberg's Theorem. The set of critical
energies is described explicitly in terms of the transmission and
reflection coefficients for scattering at the single-site
potential. In examples we discuss the asymptotic behavior of
generalized eigenfunctions at critical energies.
\end{abstract}

\maketitle

\section{The Main Result}\label{mainsec}

We study Anderson-type random Schr\"odinger operators

\begin{equation} \label{model}
H_{\omega} = -\frac{d^2}{dx^2} + \sum_{n\in\mathbb{Z}} q_n(\omega)
f(x-n) \end{equation} in $L^2(\mathbb{R})$. The {\em single site
potential} $f$ is assumed to be real, supported in $[-1/2,1/2]$,
locally in $L^1$, and not identical to $0$ (in $L^1$-sense). The
{\em coupling constants} $q_n(\omega)$, $n\in \mathbb{Z}$, are
independent, identically distributed Bernoulli random variables,
i.e.\ they have distribution $\mu$ with supp$\,\mu = \{0,1\}$.
Thus the random potential $V_{\omega}(x) = \sum_n q_n(\omega)
f(x-n)$ is in $L^1_{loc,unif}( \mathbb{R})$ for all $\omega$,
which allows to associate a unique selfadjoint operator with
$H_{\omega}$ that may equivalently be defined by form methods or
in the sense of Sturm-Liouville theory.

Our interest in Bernoulli-Anderson models of this type arises from
the fact that their spectral properties are not as well
understood, by mathematically rigorous standards, as those of
Anderson models with continuous (or absolutely continuous)
distribution $\mu$. Furthermore, these discrete distributions are
physically more relevant as they are modeling the charge numbers
of nuclei.

One of the central objects in the study of one dimensional random
operators is the {\em Lyapunov exponent}. To define it for our
model, let $g_E(n,\omega)$ be the transfer matrix of

\begin{equation} \label{diffex}
-u''+V_{\omega}u = Eu \end{equation} from $n-1/2$ to $n+1/2$,
i.e.\ for any solution of (\ref{diffex}) one has

\[ \left( \begin{array}{c} u(n+1/2) \\ u'(n+1/2) \end{array} \right) = g_E(n,\omega)
\left( \begin{array}{c} u(n-1/2) \\ u'(n-1/2) \end{array} \right).
\]

Also, for $n\in \mathbb{N}$ let $U_E(n,\omega) = g_E(n,\omega)
\ldots g_E(1,\omega)$, and define the Lyapunov exponent at $E$ by

\begin{equation} \label{lyapunov}
\gamma(E) = \lim_{n\to\infty} \frac{1}{n} \mathbb{E}(\log
\|U_E(n,\omega)\|), \end{equation} where $ \mathbb{E}$ denotes
expectation with respect to $\omega$. Existence of $\gamma(E)$
follows from the subadditive ergodic theorem, e.g.\
\cite{Carmona/Lacroix}. Since $\|U_E\|\ge 1$, we have
$\gamma(E)\ge 0$, and it can also be seen that one gets the same
$\gamma(E)$ if transfer matrices on the negative half line are
used analogously in (\ref{lyapunov}).

For many applications it is crucial to know at which energies one
has that $\gamma(E)>0$, which corresponds to exponential growth
(or decay) of the solutions to (\ref{diffex}). Our main goal here
is to show that under the above assumptions this holds for all but
a discrete set $M$ of energies $E$, and to explicitly describe $M$
in terms of the {\em transmission} and {\em reflection
coefficients} for scattering at the single site potential $f$. To
define them, let $k \in \mathbb{C}\setminus \{0\}$ and $u_+(x,k)$
be the {\em Jost solution} of

\begin{equation} \label{singlesite}
-u''+fu=k^2u, \end{equation} i.e.\ the solution satisfying

\begin{equation} \label{jostsol}
u_+(x,k) = \left\{ \begin{array}{ll} e^{ikx} & \mbox{for $x\le
-1/2$,} \\ a(k)e^{ikx} + b(k) e^{-ikx} & \mbox{for $x\ge 1/2$.}
\end{array} \right. \end{equation}
Since $e^{ikx}$ and $e^{-ikx}$ are linearly independent solutions
of $-u''=k^2u$, this defines $a(k)$ and $b(k)$ uniquely. They are
related to the transmission and reflection coefficients used in
physics by $a=t^{-1}$ and $b=rt^{-1}$. In particular, vanishing of
$b$ is equivalent to vanishing of $r$. For real $k$ we use
constancy of the Wronskian to get

\begin{equation} \label{wronsrel}
|a(k)|^2 - |b(k)|^2 = 1, \end{equation} corresponding to the
familiar $|r|^2+|t|^2=1$.

It can also be seen from (\ref{jostsol}) that

\[ \left( \begin{array}{c} a(k) \\ b(k) \end{array} \right) =
\frac{1}{2ik} e^{-ik/2} \left( \begin{array}{cc} ik & -1 \\
ike^{ik} & -e^{ik} \end{array} \right) \left( \begin{array}{c}
u_+(1/2,k) \\ u_+'(1/2,k) \end{array} \right). \] Since $u_+$ is a
solution of (\ref{singlesite}), a linear differential equation
which is analytic in $k$, and satisfies the analytic initial
condition $(u_+(-1/2), u_+'(-1/2)) = (1,ik)$, we see that
$u_+(1/2,k)$ and $u_+'(1/2,k)$ are entire in $k$. Thus $a(k)$ and
$b(k)$ are analytic in $ \mathbb{C}\setminus \{0\}$ with a
possible pole at $k=0$. Neither $a(k)$ nor $b(k)$ vanish
identically. For $a(k)$ this is a trivial consequence of
(\ref{wronsrel}). For $b(k)$ it follows from the fact that a
compactly supported $f\not=0$ can not be {\em reflectionless}
(i.e.\ a {\em soliton}), as follows from inverse scattering
theory, e.g.\ \cite{Deift/Trubowitz}, see also \cite{Sims/Stolz}.
Thus the roots of $a(k)$ and $b(k)$ can not accumulate, neither at
(the pole) $0$ nor away from $0$.

We are now ready to state our main result:

\begin{theorem} \label{mainthm} Let $E\in \mathbb{R}\setminus M$,
where the set of critical energies $M$ is given by
\begin{eqnarray} \label{critical} M & := & \{ \left(\frac{n\pi}{2}\right)^2: n\in
\mathbb{Z}\} \cup \{E=k^2: k>0, b(k)=0\} \\ & & \cup
\{-\alpha^2<0: a(i\alpha)a(-i\alpha)b(i\alpha)b(-i\alpha)=0\},
\end{eqnarray} then $\gamma(E)
>0$. \end{theorem}

Note that the critical set $M$ is the union of the discrete set
$\{ (n\pi/2)^2: n\in \mathbb{Z}\}$ and all real numbers in $\{k^2:
k\in \mathbb{C}\setminus \{0\}, a(k)=0\:\mbox{or}\:b(k)=0\}$ (this
uses that for real $k$, $a(k)\not=0$ and $b(-k)=\overline{b(k)}$).
Therefore $M$ is discrete.

Our main tool for proving Theorem~\ref{mainthm} in
Section~\ref{proofsec} will be F\"urstenberg's Theorem, which has
been used extensively in proofs of positivity of $\gamma$ for
discrete one dimensional random operators, see
\cite[Sec.14A]{Pastur/Figotin} for a summary. For continuum
models, the use of F\"urstenberg's Theorem until recently seems to
have been restricted to the special cases $f=\chi_{[-1/2,1/2]}$
and $f=\delta_0$, a $\delta$-point-interaction
\cite{Benderskii/Pastur,Ishii}. The first to have used
F\"urstenberg's Theorem systematically for continuum Anderson
models with general classes of single sites $f$ has been Kostrykin
and Schrader \cite{KS1}. While they state their general results
for absolutely continuous distribution $\mu$, they point out that
most of their ideas can also be applied to discrete distributions.
Our work here can be seen as an implementation of this fact with,
as we feel, a minimal amount of technical effort.

A nice feature of using F\"urstenberg's Theorem in the proof of
Theorem~\ref{mainthm} is that once we know that the Theorem holds
under the assumption that supp$\,\mu = \{0,1\}$, then it is
immediately clear that the Theorem extends to the case where
$\{0,1\} \subset$ supp$\,\mu$. In fact, for larger support the
critical set $M$ should be smaller. Kostrykin and Schrader note in
\cite{KS2} that $M$ should be empty if supp$\,\mu$ has at least
one non-isolated point, but there doesn't seem to be a proof of
this yet. While Theorem~\ref{mainthm} does not state that
$\gamma(E)=0$ for all $E\in M$ (which in fact is not generally
true), we will demonstrate in Section~\ref{critsec} that for
Bernoulli-Anderson models one can indeed find many critical
energies with $\gamma(E)=0$. In the discrete case this has already
been observed for the so-called {\em dimer model}, see
\cite{deBievre/Germinet}. In particular, Section~\ref{critsec}
discusses the example $f= \lambda \chi_{[-1/2,1/2]}$, $\lambda \in
\mathbb{R}$ a constant, which leads to two different types of
critical energies. They can be classified by the asymptotics of
solutions of (\ref{diffex}). In some cases they are of plane wave
type, in particular bounded, while other critical energies lead to
solutions which grow like $\exp{ \sqrt{x}}$, due to a connection
with random walks.

In \cite{DSS} we use a result much like Theorem~\ref{mainthm} to
prove exponential and dynamical localization for continuum
Bernoulli-Anderson models, where in addition we build on ideas
which were developed for discrete models in \cite{CKM}.
Discreteness of the critical set $M$ is crucial to this approach.
In particular, it shows that the result found in
Theorem~\ref{mainthm} is superior to the well known results of
Kotani theory which prove that $\gamma(E)>0$ for almost every $E$
in a class of models containing ours. The methods of \cite{DSS}
also allow to extend Theorem~\ref{mainthm} to the case where the
support of $\mu$ is $\{a,b\}$, where $a\not= b$ are arbitrary real
numbers. This requires considerably more technical effort, due to
our use of scattering theory at a periodic background, and thereby
somewhat obscures the rather simple ideas which we present here in
Section~\ref{proofsec}.

\section{Positivity of the Lyapunov Exponent}\label{proofsec}

To prove positivity of $\gamma$, we need to understand properties
of the transfer matrices. The transfer matrix from -1/2 to 1/2 of
\begin{equation} \label{sseqn}
-u^{\prime \prime} +  fu = E u,
\end{equation}
is the matrix, $g( E)$, for which
\begin{equation}
\left( \begin{array}{c} u(1/2) \\ u^{\prime}(1/2) \end{array}
\right) = g( E) \left( \begin{array}{c} u(-1/2) \\
u^{\prime}(-1/2)
\end{array} \right)
\end{equation}
for any solution $u$ of ($\ref{sseqn}$). By $g_0(E)$ we denote the
corresponding transfer matrix of $-u''=Eu$. Set $G( E )$ to be the
closed subgroup of $SL(2, \mathbb{R})$ generated by $g_0(E)$ and
$g(E)$. Let $P( \mathbb{R}^2)$ be the projective space, i.e. the
set of the directions in $\mathbb{R}^2$ and $\overline{v}$ be the
direction of $v \in \mathbb{R}^2 \setminus \{ 0 \}$. Note that
$SL(2, \mathbb{R})$ acts on $P( \mathbb{R}^2)$ by $g \overline{v}
= \overline{gv}$. We say that $G \subset SL(2, \mathbb{R})$ is
strongly irreducible if and only if there is no finite
$G$-invariant set in $P( \mathbb{R}^2)$.

In order to prove Theorem~\ref{mainthm} we will consider energies
$E>0$ and $E<0$ separately. Note that $E=0$ is contained in $M$,
so we don't need to consider it.

The general approach is as follows: We will first prove that $G(
E)$ is not compact by showing that a sequence of elements has
unbounded norm. This argument will be valid for all $E\in
\mathbb{R} \setminus M'$, where
\begin{eqnarray*}
M' & = & \{(n\pi)^2: n\in \mathbb{Z}\} \cup \{E=k^2: k>0, b(k)=0\}
\\ & & \mbox{} \cup \{ -\alpha^2<0: a(i\alpha) a(-i\alpha) b(i\alpha)
b(-i\alpha)=0\}. \end{eqnarray*} Once non-compact, then the group
$G$ is known to be strongly irreducible if and only if for each
$\overline{v} \in P(\mathbb{R}^2)$,
\begin{equation} \label{3dir}
\#\{g \overline{v}: g \in G\} \geq 3,
\end{equation}
see \cite{Bougerol/Lacroix}. In order to prove that this condition
is also satisfied we will in addition have to exclude $k$'s which
are odd multiples of $\pi/2$. We then use F\"urstenberg's Theorem
which states, in our context, that if the group $G( E)$ is not
compact and strongly irreducible then $\gamma( E)
>0$, see also \cite{Bougerol/Lacroix}.

We start with positive energies, i.e.\ $E=k^2$, $k>0$, and express
the transfer matrices over $[-1/2,1/2]$ in terms of the Jost
solutions (\ref{jostsol}), at $\pm k$. We have \[ g(E) = \left(
\begin{array}{ll} u_N(1/2,k) & u_D(1/2,k) \\ u_N'(1/2,k) & u_D'(1/2,k)
\end{array} \right) \] where $u_N$ and $u_D$ are the solutions of
(\ref{singlesite}) satisfying $u_N(-1/2,k)=u_D'(-1/2,k)$ $=1$,
$u_N'(-1/2,k)=u_D(-1/2,k)=0$. Writing $u_N$ and $u_D$ as linear
combinations of $u_+(x,k)$ and $u_-(x,k)=\overline{u_+(x,k)}$ and
setting $z_{\pm}(k) = a(k)e^{ik} \pm b(k)e^{-ik}$, we see that
\begin{equation} \label{teqn}
g(k^2) = \left( \begin{array}{cc} \mbox{Re}[z_+(k)] & \frac{1}{k}
\mbox{Im}[z_+(k)] \\ -k \mbox{Im}[z_-(k)] & \mbox{Re}[z_-(k)]
\end{array} \right).
\end{equation}

Clearly,
\begin{equation} \label{freetrans}
g_0(k^2) = \left( \begin{array}{cc} \cos k & \frac{1}{k} \sin k \\
-k \sin k & \cos k \end{array} \right). \end{equation}

Note for $k>0$,
\begin{equation} \label{conj}
\left( \begin{array}{cc} 1 & 0 \\ 0 & \frac{1}{k} \end{array}
\right) \left( \begin{array}{cc} a & \frac{1}{k}b \\ kc & d
\end{array} \right) \left( \begin{array}{cc} 1 & 0 \\ 0 & k
\end{array} \right) = \left( \begin{array}{cc} a & b \\ c & d
\end{array} \right).
\end{equation}
Thus proving non-compactness for $G(k^2)$ is equivalent to proving
non-compact\-ness for the group $\tilde{G}(k^2)$ conjugate to
$G(k^2)$ via ($\ref{conj}$).

If $b(k)=0$ we have that $|a(k)|=1$, and hence, we may write $a(k)
= e^{i \phi}$. By $( \ref{teqn})$, one has then that
\begin{equation} \nonumber
\tilde{g}(k^2) = \left( \begin{array}{cc} \cos(k+ \phi) & \sin(k +
\phi) \\ - \sin(k + \phi) & \cos(k + \phi) \end{array} \right),
\end{equation}
which is a rotation. The same holds for $g_0(k^2)$ by
(\ref{freetrans}). Thus $\tilde{G}(k^2)$ is a group of rotations
and thereby compact. This not only excludes an application of
F\"urstenberg, but shows $\gamma(k^2)=0$ as a direct consequence
of the definition (\ref{lyapunov}).

Now assume $b(k)\not=0$. Set $a(k) = Ae^{i \alpha}$, $b(k) = Be^{i
\beta}$, and see that ($\ref{wronsrel}$) implies $A^2-B^2=1$. It
follows then that $A>B>0$, $A+B>1$, and $A-B<1$. With this
notation, one may recalculate
\begin{equation} \nonumber
\tilde{g}(k^2) = A \left( \begin{array}{cc} \cos( \varphi) & -
\sin(\varphi) \\ \sin(\varphi) & \cos(\varphi) \end{array} \right)
+B \left( \begin{array}{cc} \cos(\psi) & - \sin(\psi) \\ -
\sin(\psi) & - \cos(\psi) \end{array} \right),
\end{equation}
where we have set  $\varphi:=-k- \alpha$ and $\psi:=k-\beta$.

We wish to show that a sequence of elements in $\tilde{G}(k^2)$
has unbounded norm. To do so, consider an arbitrary unit vector
\begin{equation} \nonumber
v( \theta) := \left( \begin{array}{c} \cos( \theta) \\ \sin(
\theta) \end{array} \right).
\end{equation}
The relation
\begin{equation} \nonumber
\tilde{g}(k^2)v( \theta) = A v( \varphi + \theta) + B v( - \psi -
\theta)
\end{equation}
shows that the image of the unit circle under $\tilde{g}$ is an
ellipse, with the choices $\eta:= - \frac{1}{2}( \varphi + \psi)$
and $\eta^{\prime}:= \frac{\pi}{2}- \frac{1}{2}( \varphi + \psi)$
identifying the semi-major and semi-minor axes as follows:
\begin{equation} \nonumber
\begin{array} {ccc} \tilde{g}(k^2)v( \eta)= (A+B)v( \frac{\varphi-\psi}{2})
& \mbox{and} & \tilde{g}(k^2)v( \eta^{\prime}) = (A-B)v(
\frac{\pi}{2} + \frac{\varphi-\psi}{2}) \end{array}.
\end{equation}
Defining $R( \theta):= \| \tilde{g}(k^2)v( \theta) \|$, a short
calculation, using $A^2-B^2=1$, yields
\begin{equation} \nonumber
R( \theta)-1 =2B \left[ B+A \cos( 2 \theta + \varphi + \psi)
\right].
\end{equation}
As a consequence, we see that $R( \theta)-1$ is $\pi$-periodic and
has exactly two roots in $[0,\pi)$. In particular, $R( \theta)=1$
if and only if $\cos( 2 \theta + \varphi + \psi) = - \frac{B}{A}$,
which shows that the distance between the zeros of $R( \theta)-1$
is not equal to $\frac{ \pi}{2}$: recall $B \neq 0$ (and $B \neq
A$). Similarly, $R( \theta)>1$ if and only if $\cos( 2 \theta +
\varphi + \psi) > - \frac{B}{A}$ and hence $| \{ \theta \in [0,
\pi): R( \theta)>1 \}| > \frac{ \pi}{2}$. As a result, there
exists a compact interval $K$ (not necessarily in $[0, \pi)$) with
$|K|> \frac{\pi}{2}$ and $\| \tilde{g}(k^2)v( \theta) \| > c > 1$
for all $\theta \in K$.

Now, applying $\tilde{g}$ once to $v( \eta)$, $\eta$ as above,
produces a new vector with norm greater than one, but the
direction, initially $\eta$, is possibly altered. A vector with
this new direction may not increase in norm by directly applying
$\tilde{g}$ again.  However, as long as $k$ is not an integer
multiple of $\pi$, then finitely many applications of
$\tilde{g}_0(k^2)$, which is rotation by $k$, produces a vector
with direction in $K$. Once in $K$, a direct application of
$\tilde{g}$ does increase the norm size uniformly by $c>1$, as
indicated above. In this manner, we can produce a sequence of
elements in $\tilde{G}(k^2)$ with unbounded norm. Thus
$\tilde{G}(k^2)$ is non-compact if $b(k)\not=0$ and $k$ is not an
integer multiple of $\pi$.

To complete the proof of Theorem~\ref{mainthm} for positive
energies, we note that if in addition $k$ is not an integer
multiple of $\pi/2$, then the free transfer matrix produces three
distinct elements in projective space, i.e.\ (\ref{3dir}) is
satisfied and F\"urstenberg's Theorem implies $\gamma(k^2)>0$.

\vspace{3mm}

We finally turn to energies $E<0$. In this case $G(E)$ is
non-compact for every $E < 0$, since the free transfer matrix
\begin{equation} \nonumber
g_0( E) = \left( \begin{array}{cc} \cosh( \alpha) &
\frac{1}{\alpha} \sinh( \alpha) \\ \alpha \sinh( \alpha) & \cosh(
\alpha) \end{array} \right)
\end{equation}
has unbounded powers, where $\alpha = \sqrt{| E|}$. It remains to
check ($\ref{3dir}$), which we have to do for $E \in (- \infty,0)
\setminus \{ -\alpha^2: a(i\alpha) a(-i\alpha) b(i\alpha)
b(-i\alpha)=0\}$.

Let $E=-\alpha^2$ such that $a( \pm i \alpha) \neq 0$ and $b( \pm
i \alpha) \neq 0$. By ($\ref{jostsol}$), this is equivalent to
\begin{equation} \label{unstable}
g(E) \overline{v}_{\pm} \notin \{ \overline{v}_+, \overline{v}_-
\},
\end{equation}
where $ \overline{v}_{\pm}$ are the directions of $v_{\pm} := (1,
\pm \alpha)^t$. If $\overline{v} \notin \{ \overline{v}_+,
\overline{v}_-\}$, then $\# \{ g_0( E)^n \overline{v}: n \in
\mathbb{Z} \} = \infty$. If, on the other hand, $\overline{v} \in
\{ \overline{v}_+, \overline{v}_- \}$, then we use
($\ref{unstable}$) to conclude that an initial application of $g(
E)$ followed by iteration of $g_0( E)$ gives an infinite orbit.
This shows ($\ref{3dir}$) and completes the proof of
Theorem~\ref{mainthm}.

\section{Critical energies} \label{critsec}

In this section we discuss the appearance of critical energies in
concrete examples. We thereby illustrate the following points: (i)
critical energies with vanishing Lyapunov exponent do indeed
exist, (ii) the structure and ``size'' of the set of energies with
$b(k)=0$ depends strongly on the concrete example, and (iii) at
critical energies at least two different types of non-exponential
asymptotics of solutions can be observed.

\subsection{Zero reflection at degenerate gaps} \label{nogap}

Our first general observation is that if $H^f_{{\rm per}}$ is the
Schr\"odinger operator with $1$-periodic potential $V$ satisfying
$V(x) = f(x)$, $-1/2 \le x \le 1/2$ and $H^f_{{\rm per}}$ has a
degenerate gap at energy $E = k^2$, i.e.\ either all solutions of
(\ref{singlesite}) satisfy periodic boundary conditions on
$[-1/2,1/2]$ or all solutions satisfy anti-periodic boundary
conditions on $[-1/2,1/2]$, then $g(k^2)=I$ or $g(k^2)=-I$. By
(\ref{teqn}) this implies $(a(k),b(k))=(1,0)$ or $(-1,0)$,
respectively. Conversely, $(a(k),b(k))=(\pm 1,0)$ necessarily
requires a degenerate gap. Thus, the possibility of degenerate
gaps in periodic potentials is one mechanism which leads to
reflectionless energies. Whether all reflectionless energies arise
in this way is equivalent to deciding if $b(k)=0$ necessarily
leads to $a(k)=\pm 1$. In Example 2 below we will see that this is
not the case.

\subsection{Example 1} \label{1step}
 Consider first $f := \lambda \chi_{[-1/2,1/2]}$ with
$\lambda \in \mathbb{R}$. For $E=k^2 > \max\{0,\lambda\}$ we get
with $\alpha:= \sqrt{E - \lambda}$

\begin{align*}
a(k)e^{ik} = & \, \cos \alpha + i \frac{2k^2-\lambda}{2k\alpha}
\sin \alpha \\ b(k) = & \, \frac{i \lambda}{2k\alpha} \sin
\alpha\\ g_0(k^2) = & \, \left(
\begin{array}{cc} \cos k & \frac{1}{k} \sin k \\ -k \sin k & \cos
k \end{array} \right) \\ g(k^2) = & \, \left( \begin{array}{cc}
\cos \alpha & \frac{1}{\alpha} \sin \alpha \\ -\alpha \sin \alpha
& \cos \alpha
\end{array} \right)
\end{align*}

We can now distinguish between the following types of critical
energies:

\vspace{3mm}

\noindent {\bf Type 1a:} $k^2$ such that $b(k)=0$: This happens
for $E=k^2=(n\pi)^2 + \lambda$, $n\in \mathbb{N}$. In this case we
have $g(k^2)=\pm I$, and it easily follows that the group $\langle
g_0, g \rangle$ is bounded. This means that solutions to
(\ref{diffex}) are bounded and that $\gamma(k^2)=0$.

\vspace{3mm}

\noindent {\bf Type 1b:} $k=n\pi$, $n \in \mathbb{N}$: This means
$g_0(k^2) = \pm I$, and again $\langle g_0, g \rangle$ bounded,
$\gamma(k^2)=0$. This type is ``dual'' to Type 1, since $k^2 = n^2
\pi^2$ can be viewed as the energies where $-d^2/dx^2+ \lambda -
\lambda \chi_{[-1/2,1/2]}$ is reflectionless with respect to
$-d^2/dx^2 + \lambda$.

\vspace{3mm}

\noindent {\bf Type 2:} $E=k^2$, where $k=(2n-1)\pi/2$, $n\in
\mathbb{N}$. This leads to a critical energy with a rather
different asymptotic behavior of solutions if also
$\alpha=(2m-1)\pi/2$, $m\in \mathbb{N}$, that is, for specific
values of $\lambda = k^2 - \alpha^2 = \pi^2(n-m)(n+m-1)$. In this
case we have $$ g_0(k^2) = \pm \left(
\begin{array}{cc} 0 & -1/k
\\ k & 0 \end{array} \right), \quad g(k^2) = \pm \left(
\begin{array}{cc}

0 & -1/\alpha \\ \alpha & 0 \end{array} \right). $$ Let
$h_n(\omega)
:= g_{2n+1}(\omega) g_{2n}(\omega)$, where $g_n(\omega)$ is the
transfer matrix of $H_{\omega}$ from $n-1/2$ to $n+1/2$. Then $$
h_n(\omega) = \left\{ \begin{array}{ll} \pm I & \mbox{ with
probability $p^2+q^2$, } \\ \pm \left(
\begin{array}{cc} k/\alpha & 0 \\ 0 & \alpha/k
\end{array} \right) & \mbox{ with probability $pq$, } \\ \pm \left(
\begin{array}{cc} \alpha/k & 0 \\ 0 & k/\alpha \end{array} \right) &
\mbox{ with probability $qp$. } \end{array} \right. $$

If $h_{n,1}$ and $h_{n,2}$ are the diagonal entries of $h_n$, then
the sums of $\log |h_{n,i}|$, $i=1,2$, give a symmetric random
walk (observing that $(p^2+q^2) \log 1 + pq \log (k/\alpha) + qp
\log (\alpha/k) =0$). Thus it follows that for every $\varepsilon
>0$ and almost surely
\[ \limsup_{n\to\infty} n^{-1/2-\varepsilon} \log \| \prod_{k=1}^n
h_k(\omega) \| = 0 \] and
\[ \limsup_{n\to\infty} n^{-1/2+\varepsilon} \log \| \prod_{k=1}^n
h_k(\omega) \| = \infty, \] see e.g.\ \cite{Prohorov/Rozanov}. The
interpretation of this is that the hull of the (oscillatory)
solutions of (\ref{diffex}) asymptotically grows like $\exp
c\sqrt{x}$ or decays like $\exp(-c\sqrt{x})$ for some $c>0$, where
a generalization of the Ruelle-Osceledec Theorem from
\cite[Section 8]{Last/Simon} is used.

\vspace{3mm}

\subsection{Example 2} \label{bstep}
Consider $f := \lambda (\chi_{[-1/2,0]} - \chi_{(0,1/2]})$ with
$\lambda > 0$. For $E=k^2 > \lambda$, we may again explicitly
calculate the transfer matrix $$ g(E) = \left(
\begin{array}{cc} g_{11} & g_{12} \\ g_{21} & g_{22} \end{array}
\right), $$ and see that,
\begin{align*}
g_{11} = & \, \cos (\alpha_- /2) \cos (\alpha_+ /2) -
\frac{\alpha_-}{\alpha_+} \sin (\alpha_- /2) \sin (\alpha_+ /2),
\\ g_{12} = & \, \frac{1}{\alpha_-} \sin (\alpha_- /2) \cos (
\alpha_+ /2) + \frac{1}{\alpha_+} \cos ( \alpha_- /2) \sin (
\alpha_+ /2), \\ g_{21} = & \, - \alpha_+ \cos ( \alpha_- /2) \sin
(\alpha_+ /2) - \alpha_- \sin (\alpha_- /2) \cos ( \alpha_+ /2),
\\ g_{22} = & \, - \frac{\alpha_+}{\alpha_-} \sin ( \alpha_- /2)
\sin ( \alpha_+ /2) + \cos ( \alpha_- /2) \cos ( \alpha_+ /2),
\end{align*}
for $\alpha_\pm := \sqrt{E \pm \lambda}$. From this and the
boundary conditions of the Jost solution, i.e. $$ g \left(
\begin{array}{c} e^{-ik/2} \\ ik e^{-ik/2} \end{array} \right) =
\left( \begin{array}{c} a(k) e^{ik/2} + b(k) e^{-ik/2}
\\ ik a(k) e^{ik/2} - ik b(k) e^{-ik/2} \end{array} \right), $$ one
can show that $$ b(k)=0 \mbox{ if and only if } \sin ( \alpha_+
/2) = 0 = \sin ( \alpha_- /2). $$

Therefore, if we fix some $\lambda
> 0$, then any $E = k^2 > \lambda$ for which $b(k) = 0$ must
satisfy $$ \frac{\sqrt{E+\lambda}}{2} = n \pi, \mbox{ and }
\frac{\sqrt{E-\lambda}}{2} = m \pi, $$ for some $n,m \in
\mathbb{N}$. Thus $\lambda$ must have the form
\begin{equation}\label{lnec}
\lambda = 2 \pi^2 (n^2 - m^2), \mbox{ with } n,m \in \mathbb{N},
\end{equation}
and the corresponding reflectionless $E$'s are then
\begin{equation}\label{Enec}
E = 2 \pi^2 (n^2 + m^2).
\end{equation}

From this we conclude that for most values of $\lambda$ (e.g.,
Lebesgue almost every $\lambda$), no reflectionless energy exists
above $\lambda$. On the other hand, every $\lambda$ satisfying
\eqref{lnec} has only finitely many reflectionless energies. In
fact, if $N := \frac{\lambda}{2 \pi^2} \in \mathbb{N}$, then $N =
n^2 - m^2$ for only finitely many pairs $(n,m) \in \mathbb{N}^2$.
But this finite number can be arbitrarily large, depending on
$\lambda$: To find at least $j$ exceptional energies, choose $$
\frac{ \lambda_j}{2 \pi^2} = N_j = 2^{j+1}(2^{j-1}+1) $$ and see
that for $0 \leq \ell < j$ one may choose $$ n_{\ell} := \left(
2^{\ell}(2^{j-1}+1)+2^{j-\ell-1} \right) $$ and $$ m_{\ell} :=
\left( 2^{\ell}(2^{j-1}+1)-2^{j-\ell-1} \right) $$ satisfying
$n_{\ell}^2-m_{\ell}^2= N_j$. Proof: Take $a:=2^{\ell}(2^{j-1}+1)$
and $b:=2^{j-\ell-1}$ and see that $(a+b)^2-(a-b)^2=4ab$.
Therefore, $$ 4ab=2^2 \cdot 2^{\ell}(2^{j-1}+1) \cdot 2^{j-\ell-1}
= 2^{j+1}(2^{j-1}+1)= N_j. $$

Recall from our remarks in \ref{nogap} that in some cases, the
condition $b(k)=0$ implies that $a(k) = \pm 1$. To see that this
is not always the case, we calculate $a(k)$ in this example for
such values of $k$. For $\lambda$ as in \eqref{lnec} and $E$ as in
\eqref{Enec}, one can show that $a(k) = e^{-2 \pi i \sqrt{n^2 +
m^2}}$ if $n+m$ is even and $a(k) = - e^{-2 \pi i \sqrt{n^2 +
m^2}}$ if $n+m$ is odd.


\begin{thebibliography}{99}


\bibitem{Benderskii/Pastur} M.\ Benderskii and L.\ Pastur, On the asymptotics of
the solutions of second-order equations with random coefficents
(in Russian). \textit{Teoria Funkcii, Func.\ Anal.\ i Priloz.}
(Kharkov University) {\bf N 22} (1975), 3--14

\bibitem{deBievre/Germinet} S.\ de Bi\`{e}vre and F.\ Germinet, Dynamical Localization
for the Random Dimer Schr\"{o}dinger Operator. \textit{J.\ Stat.\ Phys.} {\bf 98}
(2000), 1135--1148

\bibitem{Bougerol/Lacroix} P.\ Bougerol and J.\ Lacroix, \textit{Products of random matrices
with applications to Schr\"odinger operators}. Birkh\"auser,
Boston--Stuttgart (1985)

\bibitem{CKM} R.\ Carmona, A.\ Klein, and F.\ Martinelli, Anderson localization for
Bernoulli and other singular potentials. \textit{Commun.\ Math.\ Phys.} {\bf 108}
(1987), 41--66

\bibitem{Carmona/Lacroix} R. Carmona and J. Lacroix, \textit{Spectral theory of
random Schr\"odinger operators}. Birkh\"auser, Basel--Berlin (1990)

\bibitem{DSS} D.\ Damanik, R.\ Sims and G.\ Stolz, Localization for one dimensional,
continuum, Bernoulli-Anderson models. Preprint 2000, mp-arc/00-404

\bibitem{Deift/Trubowitz} P.\ Deift and E.\ Trubowitz, Inverse scattering
on the line. \textit{Commun.\ Pure Appl.\ Math.} {\bf 32} (1979),
121--251

\bibitem{Ishii} K.\ Ishii, Localization of eigenstates and transport phenomena in
one-dimensional disordered systems. \textit{Progress Theor.\
Phys.\ Suppl.} {\bf 53} (1973), 77--118

\bibitem{KS1} V.\ Kostrykin and R.\ Schrader, Scattering theory approach to random Schr\"odinger
operators in one-dimension. \textit{Rev.\ Math.\ Phys.} {\bf 11}
(1999), 187--242

\bibitem{KS2} V.\ Kostrykin and R.\ Schrader, Global bounds for the Lyapunov exponent and
the integrated density of states of random Schr\"odinger operators
in one dimension. Preprint 2000, mp-arc/00-226

\bibitem{Last/Simon} Y.\ Last and B.\ Simon, Eigenfunctions,
transfer matrixes, and absolutely continuous spectrum of
one-dimensional Schr\"odinger operators. \textit{Invent.\ math.}
{\bf 135} (1999), 329--367

\bibitem{Pastur/Figotin} L.\ Pastur and A.\ Figotin, \textit{Spectra of Random and
Almost-Periodic Operators}. Springer Verlag, Berlin-Heidelberg-New
York (1992)

\bibitem{Prohorov/Rozanov} Yu.\ V.\ Prohorov and Yu.\ A.\ Rozanov, \textit{Probability
Theory}. Springer Verlag, Berlin-Heidelberg-New York (1969)

\bibitem{Sims/Stolz} R.\ Sims and G.\ Stolz, Localization in one dimensional random
media: a scattering theoretic approach. \textit{Commun.\ Math.\
Phys.} {\bf 213} (2000), 575--597


\end{thebibliography}
\end{document}